\documentclass{article}

    \usepackage[final]{neurips_2023}

\usepackage[utf8]{inputenc} 
\usepackage[T1]{fontenc}    
\usepackage{hyperref}       
\usepackage{url}            
\usepackage{booktabs}       
\usepackage{amsfonts}       
\usepackage{nicefrac}       
\usepackage{microtype}      
\usepackage{xcolor}         
\usepackage{graphicx}

\title{Is 3-(F)WL Enough to Distinguish All 3D Graphs?}

\author{%
  Wanghan Xu \\
  Xi’an Jiaotong University \\
  \texttt{WhoisYt@stu.xjtu.edu.cn} \\
}

\begin{document}

\maketitle

\begin{abstract}
The problem of graph isomorphism is an important but challenging problem in the field of graph analysis, for example: analyzing the similarity of two chemical molecules, or studying the expressive ability of graph neural networks. WL test is a method to judge whether two graphs are isomorphic, but it cannot distinguish all non-isomorphic graphs. As an improvement of WL, k-WL has stronger isomorphism discrimination ability, and as k increases, its discrimination ability is strictly increasing. However, whether the isomorphic discrimination power of k-WL is strictly increasing for more complex 3D graphs, or whether there exists k that can discriminate all 3D graphs, remains unexplored. This paper attempts to explore this problem from the perspective of graph generation.
\end{abstract}

\section{WL test and graph generation}

The process of WL test is to map graphs to labels. When the the multiset of the final labels of two graphs are the same, but the two graphs are not isomorphic, we call these two graphs a counterexample. If we take the WL test as a function $WL(\cdot)$, then the input of its inverse function is a multiset of labels, and its output is a graph. We can regard the inverse function of WL test as a graph generation function $GG(\cdot)$. Exploring the corresponding relationship between the input and output of this function can help us analyze whether there are counterexamples in WL test. In extreme cases, if $GG(\cdot)$ is one to one (one input to one output), there must be no counterexample in this WL test. But if $GG(\cdot)$ is one to many, we also hope that the number of outputs is as small as possible. Because the fewer graphs a set of labels can generate through $GG(\cdot)$, the more spatial information is contained in these set of labels.

\section{Tricks in graph generation}
For a WL test algorithm and a set of labels, there are often many generated graphs, which is due to three uncertainties in the generation process, that is, \textbf{exchange tricks}, \textbf{turn-over tricks} and \textbf{symmetry tricks}.

\subsection{Exchange tricks}
\begin{figure}[htbp]
  \centering
  {\includegraphics[width=11cm]{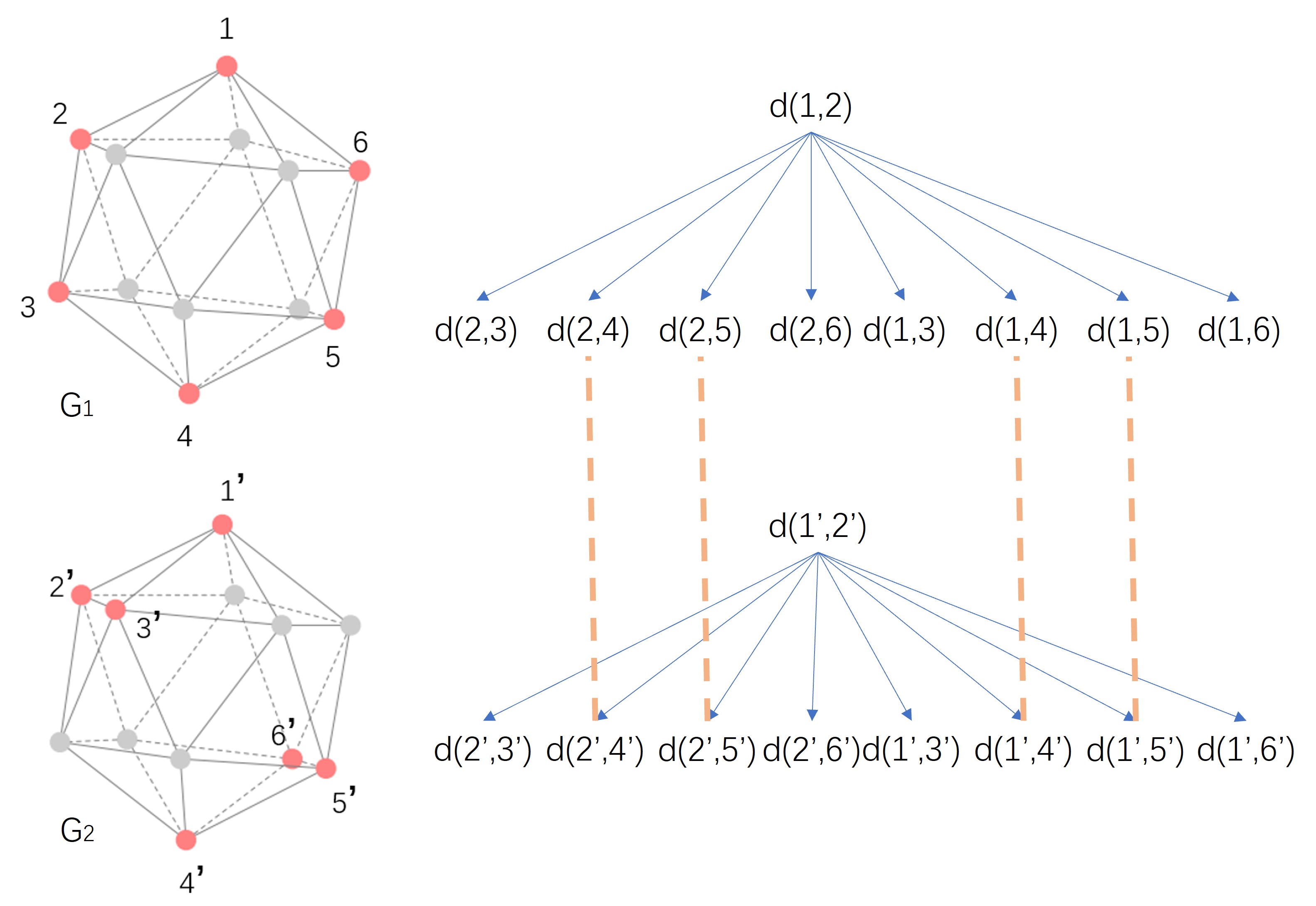}}
  \label{fig: Counterexample of 2-WL and WL tree of 2-tuple.}
  \caption{Counterexample of 2-WL and WL tree of 2-tuple. Yellow dotted lines indicate equal values.}
\end{figure}

Let us first understand A through a concrete example. Figure \ref{fig: Counterexample of 2-WL and WL tree of 2-tuple.} shows a group of pictures that cannot be  discriminated by 2-WL. By comparing $G_1$ and $G_2$, it is easy to find that the difference between them is that node $3$ in $G_1$ is changed to node $3'$ in $G_2$, and node $6$ in $G_1$ is changed to node $6'$ in $G_2$. But why can't 2-WL catch such a change? 

For 2-WL, the $N_1$ neighbor of the 2-tuple $(i, j)$ is to replace i with any node, that is, the set of edges from all nodes to $j$. Similarly, the $N_2$ neighbors of $(i, j)$ are the set of edges from all nodes to $i$. Initialize 2-tuple with distance ($d(i,j)$), then the two-layer WL tree of the 2-tuple $(1, 2)$ and the 2-tuple $(1', 2')$ is as shown on the right side of Figure \ref{fig: Counterexample of 2-WL and WL tree of 2-tuple.}. 

It can be seen from the WL tree that node $3$ is associated with edge $(1, 2)$ through two distances, that is $d(1, 3)$ and $d(2, 3)$. Suppose $d(1, 3)$ is $d_1$, and $d(2 ,3)$ is $d_2$. Similarly, node $6$ is associated with edge (1,2) through $d(1, 6)$ and $d(2, 6)$. Suppose $d(1, 6)$ is $d_3$, and $d(2, 6)$ is $d_4$. We call $\{d_1, d_2\}$ the associated information $inf((1, 2), 3)$ of node $3$ on edge $(1, 2)$. Similarly, $inf((1, 2), 6) = \{d_3, d_4\}$.

\begin{figure}[htbp]
  \centering
  {\includegraphics[width=11cm]{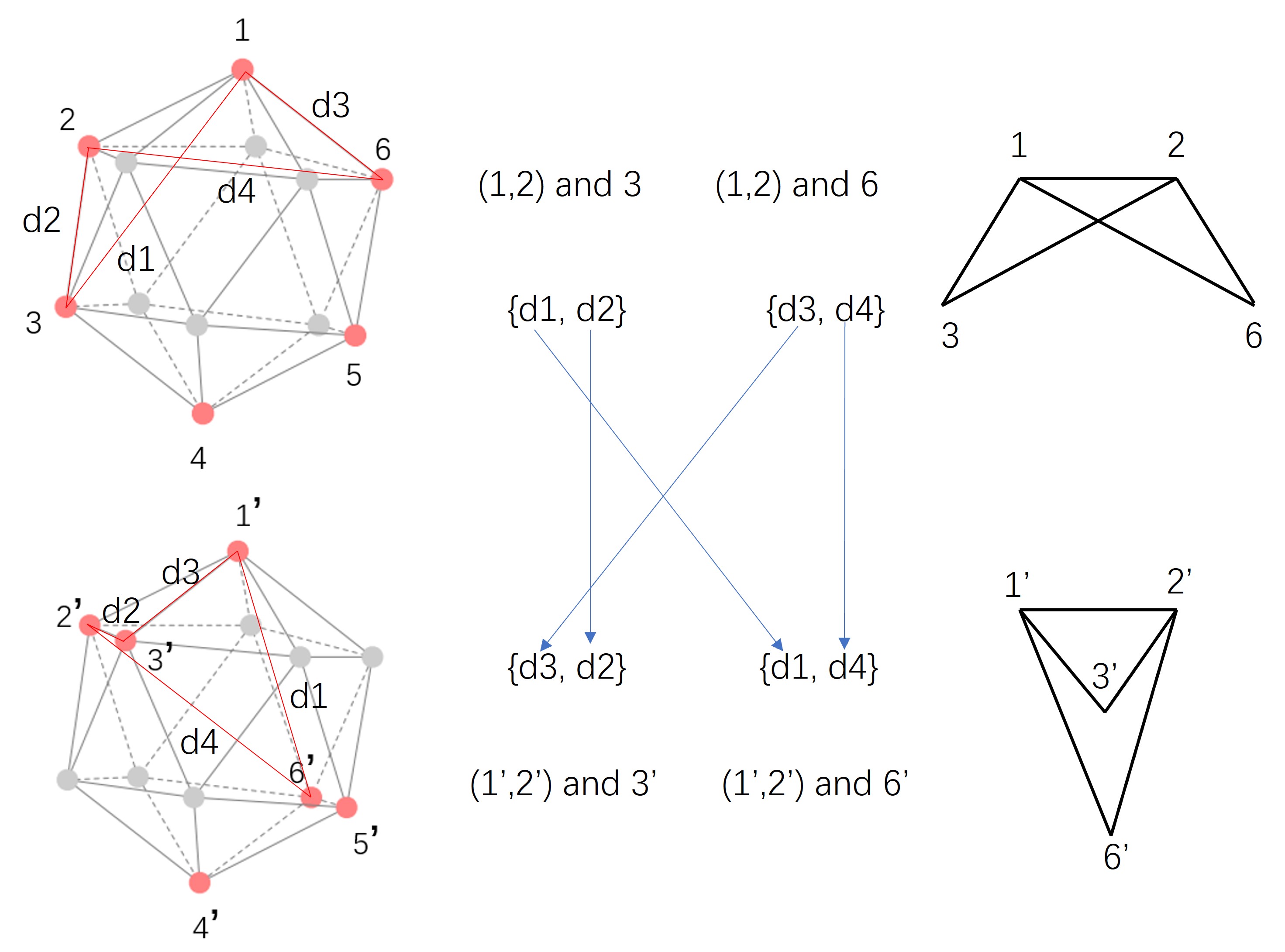}}
  \label{fig: Exchange trick.}
  \caption{Exchange trick.}
\end{figure}

In the reverse process, graph generation, of a WL test algorithm, although we can know the WL tree of tuple $(1, 2)$ based on the tuple label and the update function of WL test, we cannot divide $d_1, d_2, d_3, d_4$ into two groups and figure out $inf((1, 2), 3) = \{d_1, d_2\}$ or $inf((1, 2), 6) = \{d_3, d_4\}$. In fact, by exchanging $d_1$ and $d_3$ in $inf((1, 2), 3)$ and $inf((1, 2), 6)$, we can generate new associated information $inf((1 , 2), 3') = \{d_3, d_2\}$ and $inf((1, 2), 6') = \{d_1, d_4\}$, which is also geometrically true. Figure \ref{fig: Exchange trick.} shows that the same set of labels can correspond to different graphs due to this exchange tricks.

Formally, for a k-tuple $p$ and nodes $i, j$ that do not belong to $p$, when the graph generation process cannot distinguish whether a $p$'s neighbor $q$ belongs to $inf(p, i)$ or $inf(p, j)$, then some elements in $inf(p, i)$ and $inf(p, j)$ can be exchanged, resulting in non-unique output graphs. We call this uncertainty the \textbf{exchange trick}.

\subsection{Turn-over tricks}
Besides the exchange trick, another type of uncertainty that arises during graph generation is the turn-over trick, which means that even if the associated information is not exchanged, there is still uncertainty when the elements belonging to a set of associated information are combined. 

\begin{figure}[htbp]
  \centering
  {\includegraphics[width=9cm]{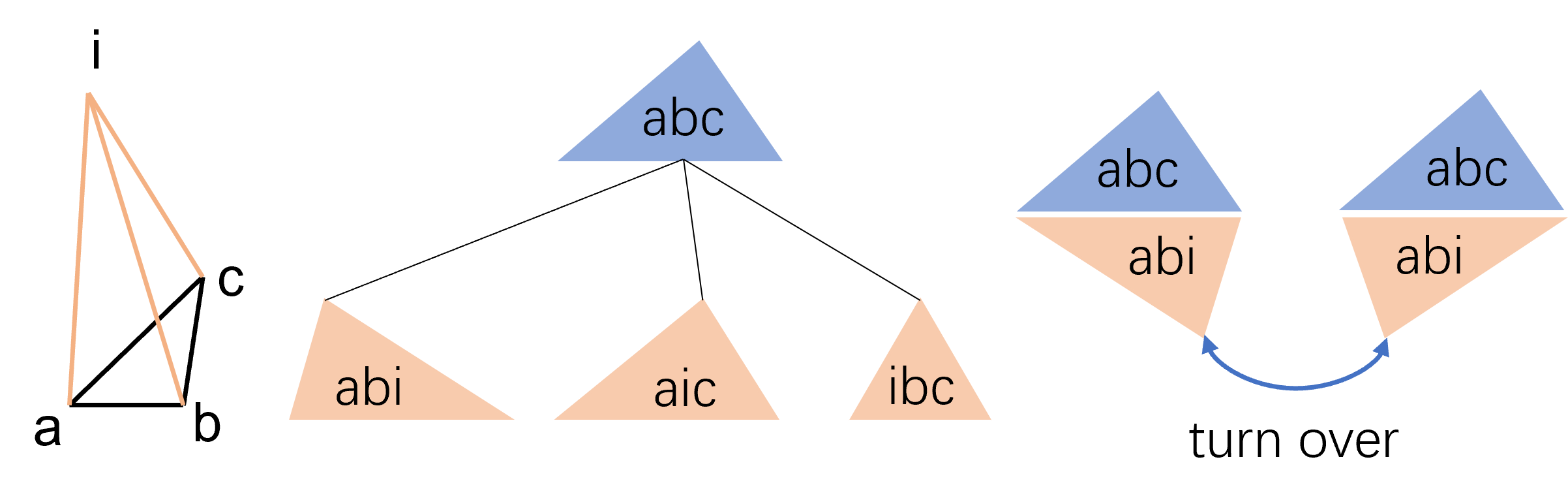}}
  \label{fig: Turn-over trick.}
  \caption{Turn-over trick.}
\end{figure}

As shown in Figure \ref{fig: Turn-over trick.}, suppose $inf((a, b, c), i) = \{(a, b, i), (a, i, c), (i, b, c)\}$, but when we specifically analyze the connection relationship between these 3-tuples (triangles), we cannot determine the direction of them. For example, for $(a, b, i)$, it has a common edge with $(a, b, c)$ (the method of determining the common edge will be introduced later), but by turning over, $(a, b, c)$ and $(a, b, i)$ have two ways of connection, which leads to non-unique output graphs.

\subsection{Symmetry tricks}
\begin{figure}[htbp]
  \centering
  {\includegraphics[width=2cm]{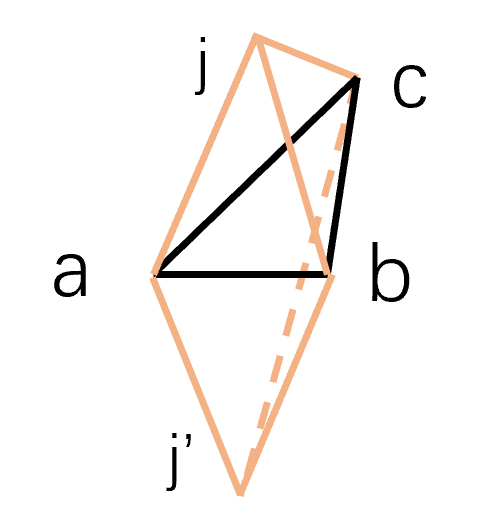}}
  \label{fig: Symmetry trick.}
  \caption{Symmetry trick.}
\end{figure}

The symmetry trick is that even though the exchange tricks and the turn-over tricks do not occur during graph generation, neighbor k-tuples still have multiple possible spatial positions that are symmetric about the root k-tuple. As shown in Figure \ref{fig: Symmetry trick.}, the spatial relationship of $j$ with respect to $(a, b, c)$ has been determined, but there are still spatial positions in two spaces that are symmetrical about $(a, b, c)$. 

\section{3-FWL doesn't have tricks}
In this section, we will start from the update function \ref{con: 3-FWL} of 3-FWL and prove that during the graph generation process of 3-FWL, there will be no exchange tricks, turn-over tricks or symmetry tricks. 

\begin{equation}
c^{t+1}_v=HASH(c^t_v,\{\{(c^t_{\Phi_1(v,j)},c^t_{\Phi_2(v,j)},c^t_{\Phi_3(v,j)})|j\in[N]\}\})
\label{con: 3-FWL}
\end{equation}

\subsection{3-tuple initialization}
When initializing the color of 3-tuples, it is necessary to satisfy "same color $\iff $ node isomorphism". There are many initialization methods that meet the conditions. The following is a color initialization method we designed (for the convenience of subsequent analysis, this initialization method is used by default in this paper):

For 3-tuples $(a, b, c)$, $d(a, b)$ represents the distance from node $a$ to node $b$ (if $a = b$, then $d = 0$). Arrange $d(a, b), d(b, c), d(c, a)$ from small to large to form a new tuple $(d_1, d_2, d_3)$. Corresponding $d_1, d_2$, and $d_3$ to the values of the three channels in the RGB color respectively, the initial color $c^0 = (d_1, d_2, d_3)$ of $(a, b, c)$ is obtained. (In detail, in order to prevent the distance from exceeding the upper limit 255 of RGB, all distances can be normalized.)

RGB color initialization meets the initialization requirements, because for two 3-tuples, if the multiple sets of distances between them are the same, i.e. $(d_1, d_2, d_3)$ = $(d_{1}', d_{2}', d_{3}')$, then the two tuples must be isomorphic. The purpose of choosing this initialization method is that the lengths of the three edges in the 3-tuples can be easily obtained through the color.

\subsection{3-FWL doesn't have exchange tricks}
It can be seen from function \ref{con: 3-FWL} that $\{c^t_{\Phi_1(v,j)},c^t_{\Phi_2(v,j)},c^t_{\Phi_3(v,j)}\} = inf(v, j)$. In other words, 3-FWL does not need to group tuples to determine which tuples belong to the associated information of the same node. This is a natural advantage of 3-FWL. 

\subsection{Identify common edges}
\label{sec: Identify common edges.}
Select any 3-tuple $v$ without repeated nodes, and set $v = (a, b, c)$. In function \ref{con: 3-FWL}, $c^t_{\Phi_1(v,j)}$ means to replace the first element $a$ in 3-tuple $v$ with $j$. After initialization with RGB, $c^0_{\Phi_1(v,j)}$ can be expressed as $(d_{j,1}, d_{j,2}, d_{j,3})$. When $j$ traverses all nodes, $(d_{j,1}, d_{j,2}, d_{j,3})$ can be divided into two types:

\begin{itemize}
\item 1) $\ 0 < d_{j,1} \le d_{j,2} \le d_{j,3}$, when $j \notin \{b, c\}$.
\item 2) $\ 0 = d_{j,1} < d_{j,2} = d_{j,3} = d(b, c)$, when $j \in \{b, c\}$.
\end{itemize}

Pick out all $c^0_{\Phi_1(v,j)}$ belonging to class 2), two in total, that is $j = b$ or $j = c$. Then we can know that the length of the common edge between $\Phi_1(v,j)$ and $v$ is $d(b, c)$, denoted as $CE(\Phi_1(v,j), v) = d(b, c)$. Although we don't know the specific numbers of $a, b, c$, we can know which edge of the 3-tuple $\Phi_1(v,j)$ is the common edge in $v$ through $CE(\Phi_1(v,j), v) = d(b, c)$. In the same way, we can also identify the common edge of $\Phi_2(v,j)$ and $v$, and the common edge of $\Phi_3(v,j)$ and $v$.

After knowing the common edges between $\Phi_1(v,j),\Phi_2(v,j),\Phi_3(v,j)$ and $v$, we are equivalent to knowing how $\Phi_1(v,j),\Phi_2(v,j),\Phi_3(v,j)$ and $v$ are connected.

\subsection{3-FWL doesn't have turn-over tricks}
\label{sec: turn-over trick.}

After identifying common edges, $\Phi_i(v,j)$ and $v$ are connected by $CE(\Phi_1(v,j), v)$, but there is still an uncertainty, that is, turn-over tricks. Except the common edge, we call the other two edges in $\Phi_1(v,j)$ the new edges (NE). Then after removing $CE(\Phi_1(v,j), v)$, $c_{\Phi_1(v,j)}$ becomes $(NE_1, NE_2)$ from $(d_{j,1}, d_{j,2}, d_{j,3})$. Do the same for $c_{\Phi_2(v,j)}$ and $c_{\Phi_2(v,j)}$ to get $(NE_3, NE_4)$ and $(NE_5, NE_6)$. 

These new edges form a multiset $NE_{SET}=\{\{NE_1, NE_2, NE_3, NE_4, NE_5, NE_6\}\}$, which is actually the set of three edges copied once (since each edge appears in two tuples, as shown in Figure \ref{fig: Turn-over trick.}). The histogram of $NE_{SET}$ has only the following three possible situations:

\begin{itemize}
\item 1) Three lengths, and the ratio is 2:2:2. That is, the three edges are not equal to each other.
\item 2) Two lengths, and the ratio is 4:2. That is, two of the three edges are equal.
\item 3) One length. That is, all three edges are equal.
\end{itemize}

\begin{figure}[htbp]
  \centering
  {\includegraphics[width=4.5cm]{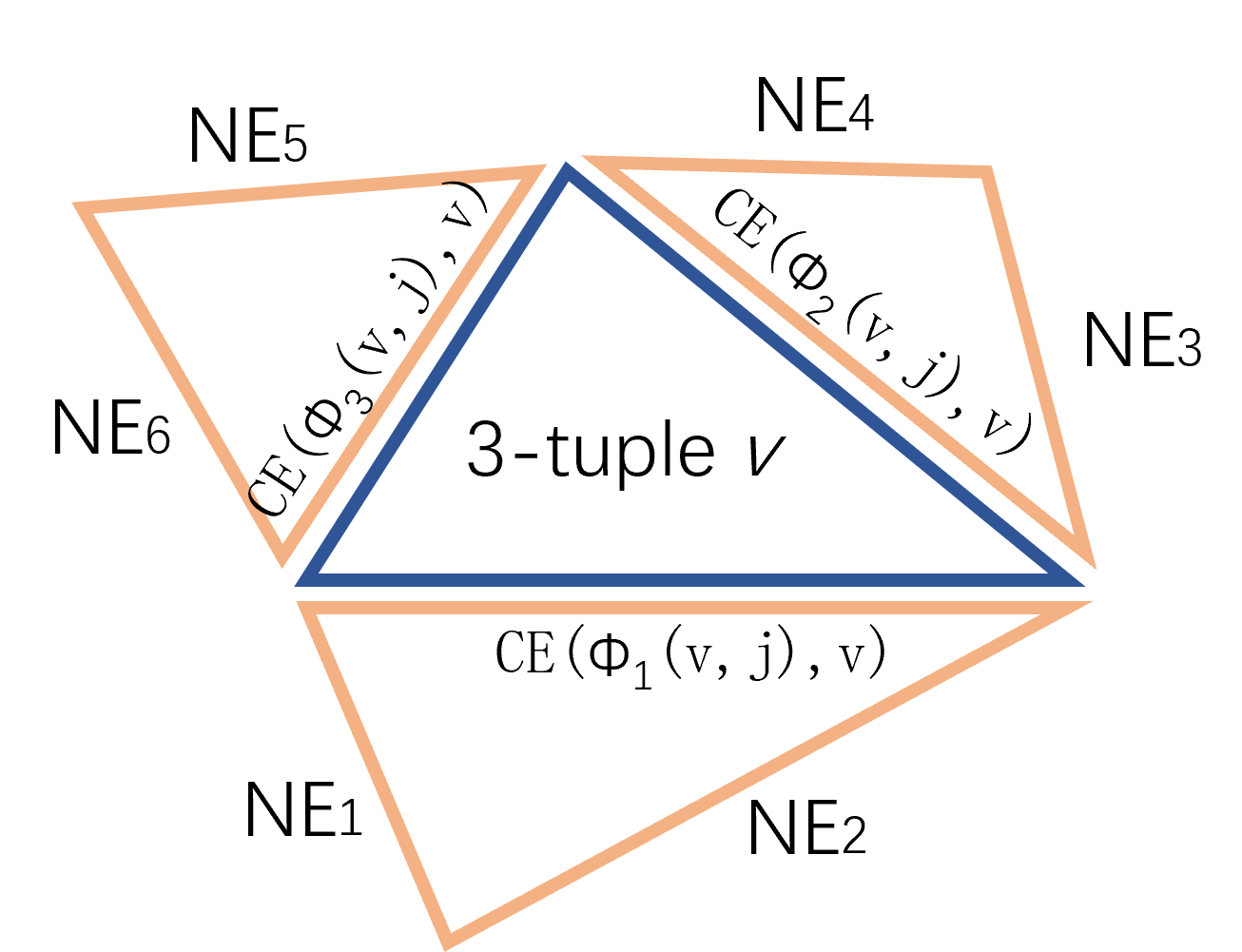}}
  \label{fig: NE.}
  \caption{Examples of new edges (NE) and $v$.}
\end{figure}

For case 1), turn-over tricks cannot happen. As shown in the Figure \ref{fig: NE.}, the premise of $NE_1$ and $NE_2$ turn-over trick is: ($NE_1=NE_4$ OR $NE_1=NE_3$) AND ($NE_2=NE_5$ OR $NE_2=NE_6$), which is contradictory to case 1). In the same way, $NE_3$ and $NE_4$ turn-over trick, and $NE_5$ and $NE_6$ turn-over trick are also impossible to happen.

For case 2), turn-over tricks may occur. But these turn-over tricks don't change any spatial relationships. Assume that among the three edges, the edges' lengths corresponding to $NE_1$ ($NE_6$) and $NE_2$ ($NE_3$) are equal, that is, ($NE_1=NE_2=NE_3=NE_6$) AND ($NE_4=NE_5$) AND ($NE_1!=NE_4$). At this time, it is impossible for $NE_5$ and $NE_6$ to turn over because $NE_5$ is not equal to $NE_1$ or $NE_2$. Similarly, it is impossible for $NE_3$ and $NE_4$ to turn over. $NE_1$ and $NE_2$ may have turn-over trick, but this trick will not affect any spatial relationship, because $NE_1=NE_2$, the spatial structure remains unchanged before and after turning over.

For case 3), turn-over tricks may occur, but these tricks don't change any spatial relationship. Because all $NE$s have the same length, the spatial structure remains unchanged before and after turning over.

In summary, for any $(c^t_{\Phi_1(v,j)},c^t_{\Phi_2(v,j)},c^t_{\Phi_3(v,j)})$ in the function \ref{con: 3-FWL}, it is impossible for them to turn over when they are spliced with the root 3-tuple $v$. After determining the connection shape between the $\Phi_i(v,j),i=1,2,3$ and $v$, "close" the three faces $\Phi_i(v,j),i=1,2,3$ to obtain a tetrahedron with $(a, b, c)$ as the base and $j$ as the upper vertex. In turn, the spatial relationship of node $j$ to $(a, b, c)$ can be obtained.

\subsection{3-FWL doesn't have symmetry tricks}

Although the relative position of $j$ with respect to $(a, b, c)$ is deterministic, it is still uncertain which edge of the face $j$ is on, that is symmetry trick, as shown in Figure \ref{fig: Symmetry trick.}. 

\begin{figure}[htbp]
  \centering
  {\includegraphics[width=9cm]{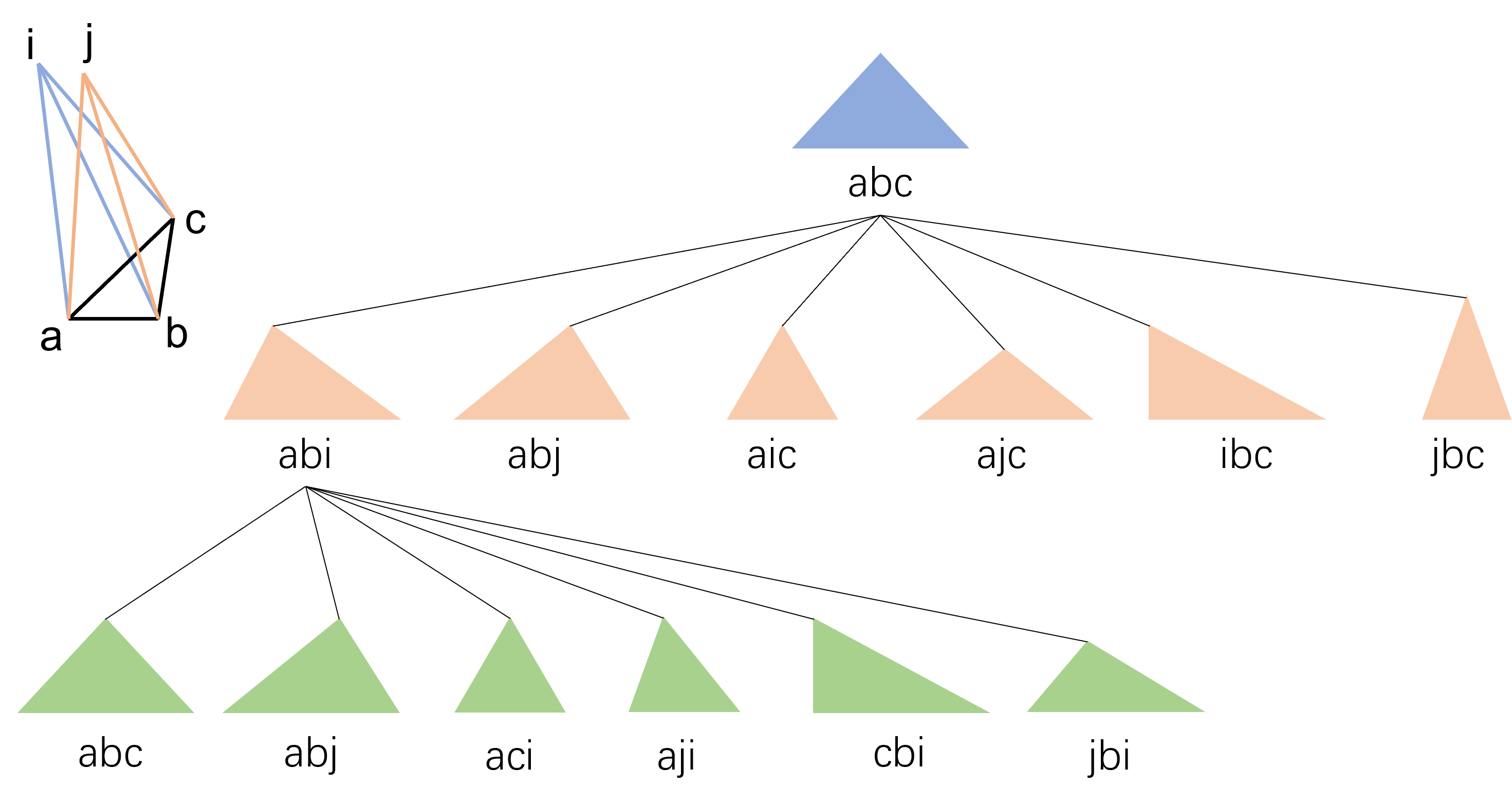}}
  \label{fig: 3-layer WL tree.}
  \caption{3-layer WL tree.}
\end{figure}

The above analysis only uses the information of the 2-layer WL tree, which can help determine the spatial relationship between "root tuple and neighbors", but cannot determine any spatial relationship between "neighbors and neighbors". In order to further determine the spatial relationship between neighbors, neighbors' neighbors must be expanded, that is, a 3-layer WL tree is used, as shown in Figure \ref{fig: 3-layer WL tree.}.

We call the possible spatial positions of all external nodes (nodes other than those in $v$) of the root tuple $(a, b, c)$ as candidate point $(CP)$. Because the spatial positions of external nodes cannot coincide, and their symmetry positions about a plane cannot coincide, so $|CP|=2(n-3)$ (Suppose the number of nodes in the graph is n). It should be noted that when a graph is determined, if all external nodes of $(a, b, c)$ are symmetrical about the $(a, b, c)$ plane, then a mirrored and same graph will be generated. In order to avoid this situation, we need to randomly select a node among the external nodes, and fix its position first, so as to distinguish the two edges of $(a, b, c)$. This node is called anchor, denoted as $m$. Remove it and its symmetry points from the $CP$ set. Therefore, we use $CP((a, b, c), m)$ to describe the candidate point set. After removing the anchor point and its symmetry point, $|CP((a, b, c), m)|=2(n-4)$. When $n=5$, the example of $CP$ is as follows:

\begin{figure}[htbp]
  \centering
  {\includegraphics[width=2cm]{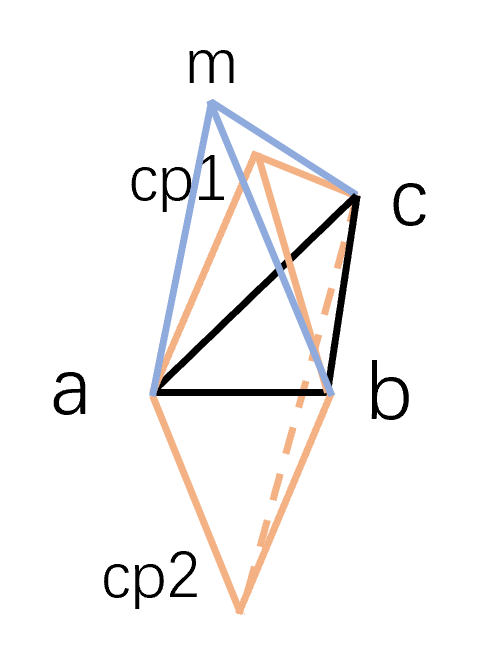}}
  \label{fig: Candidate point.}
  \caption{Candidate point.}
\end{figure}

Now, we pick a neighbor $(a, b, m)$ of the root tuple and continue to expand its neighbors. The process of analyzing "neighbors of neighbors" is exactly the same as the process of analyzing "neighbors of root nodes". This means that we can also generate a candidate point set `$CP((a, b, m), c)$ of $(a, b, m)$. Let's look at the association of $CP((a, b, c), m)$ and $CP((a, b, m), c)$.

$CP((a, b, c), m)$ and $CP((a, b, m), c)$ both describe $(V-{a, b, c, m})$ (suppose $V$ is the set of all nodes) a total of $(n-4)$ nodes $2\times(n-4)$ possible spatial positions. It can be seen from the existence of the graph that there are at least $(n-4)$ same points in $CP((a, b, c), m)$ and $CP((a, b, m), c)$ (for real points, It must appear in every candidate point set). When the number of the same points (intersection) in $CP((a, b, c), m)$ and $CP((a, b, m), c)$ is $(n-4)$, the spatial positions of all nodes are uniquely identified. In fact, $|CP((a, b, c), m) \cap  CP((a, b, m), c)|$ may be larger than $(n-4)$. Still, we want this number to be as small as possible, because it means we can determine the positions of more points.

\begin{figure}[htbp]
  \centering
  {\includegraphics[width=7cm]{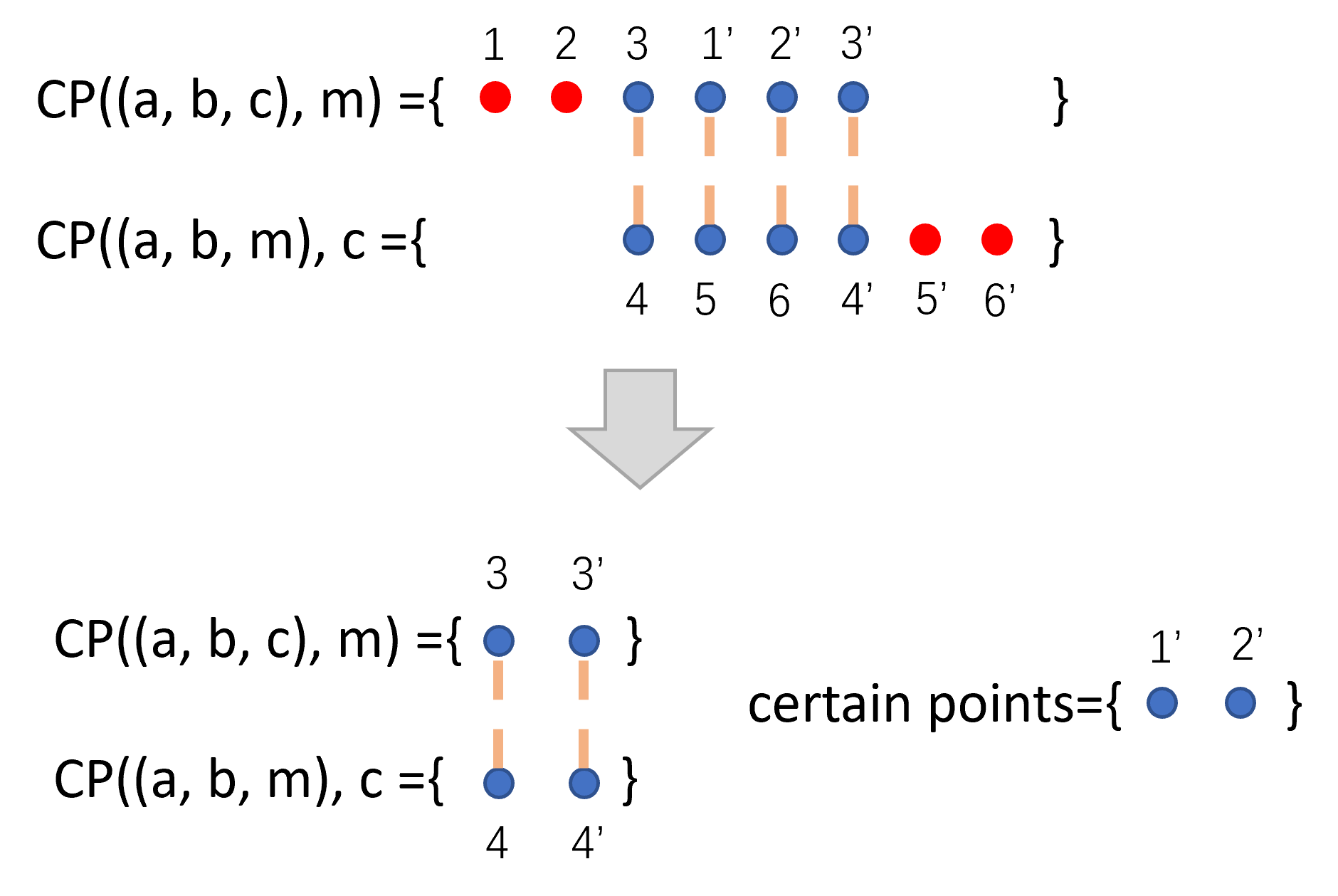}}
  \label{fig: Determine node positions.}
  \caption{Determine node positions by comparing different candidate point sets. }
\end{figure}

Each point in the set represents a spatial position where a node may exist. 1 and 1' are points of symmetry about the plane $(a,b,c)$, and so on. Blue represents the parts that are the same in both sets, red is the remainder. By comparing the two sets, it can be concluded that there is no node in the red candidate point. This is because it does not exist in another candidate point set, that is, for another plane, it is impossible to have a point at this spatial position. Therefore, red candidate points should be removed. When 1 is removed, there must be a node for 1', so add 1' to the certain points set. After removing the points that cannot exist and the points that have been determined, only 3 and 3' are left in the candidate set, which are still undetermined. After the candidate points are deleted by the above method, the remaining candidate points must be shared by all $CP$s and symmetrical to the plane of each $CP$.

If the intersection of multiple $CP$s is not an empty set, there are always undetermined nodes. But in reality, this is unlikely to happen. Suppose there is a subset $W$ of a $CP$, $W$ is the set of candidate points that cannot determine through $CP_1$, $CP_2$, $CP_3$, $CP_4$ (respectively referring to $CP((a, b, c), m)$, $CP((a, b, m), c)$, $CP ((a, m, c), b)$, $CP((m, b, c), a)$). From Figure \ref{fig: Determine node positions.}, it can be seen that $W$ must meet the following two conditions:

\begin{itemize}
\item 1) Every node in $W$ appears in $CP_1, CP_2, CP_3, CP_4$.
\item 2) The points in $W$ are symmetrical about $(a, b, c)$, $(a, b, m)$, $(a, m, c)$, $(m, b, c)$.
\end{itemize}

In physics, the symmetry plane of a particle system must pass through its center of gravity. It is easy to conclude that each symmetry plane in $W$ must intersect at one point (which is the center of gravity after taking each point in $W$ as a mass point with the same mass). 

However, since $(a, b, c), (a, b, m), (a, m, c), (m, b, c)$ are four faces of the tetrahedron $(a, b, c, m)$, they certainly not at one point. Therefore, $W$ is asymmetric about at least one of the planes. This contradicts condition 2), for which $W$ does not exist. That is, after comparing $CP_1, CP_2, CP_3, CP_4$, it is impossible for any node to remain symmetrical, and it is impossible for any node to remain undetermined.

\subsection{3-FWL can generate a unique graph}

According to the 3-FWL algorithm (including HASH correspondence) and the label of the tuple, we can get the WL tree of the tuple. 

Since there is no exchange trick in 3-FWL, each $(c^t_{\Phi_1(v,j)},c^t_{\Phi_2(v,j)},c^t_{\Phi_3(v,j)})$ in function \ref{con: 3-FWL} corresponds to the associated information of an external node. 

After identifying the common edges, triangles $\Phi_i(v,j),i=1,2,3$ can be spliced with triangle $v$. And since there is no turn-over trick in 3-FWL, there is only one correct way of splicing.

After the splicing, the spatial relationship of the external nodes with respect to the bottom face $(a, b, c)$ can be obtained by closing the triangles. Since 3-FWL does not have symmetry tricks, it can only generate one graph in the end.

\subsection{3-FWL can distinguish all 3D graphs}
Given the 3-FWL algorithm, a set of node labels generates a unique 3D graph. The converse-negative proposition is that two non-isomorphism 3D graphs generates different labels through 3-FWL. Therefore, there is no counterexample to 3-FWL, that is, 3-FWL can distinguish all 3D graphs.

\section{Does 3-WL have tricks?}
The update function of 3-WL is as follows:

\begin{equation}
c^{t+1}_v=HASH(c^t_v,\{\{c^t_{\Phi_1(v,j)}|j\in[N]\}\},\{\{c^t_{\Phi_2(v,j)}|j\in[N]\}\},\{\{c^t_{\Phi_3(v,j)}|j\in[N]\}\})
\label{con: 3-WL}
\end{equation}

By comparing with function \ref{con: 3-FWL}, we can find that the biggest difference between them is that the update function of 3-WL cannot directly obtain which $\Phi_i(v,j)$s are the associated information belonging to the same node. In other words, it is difficult to find $\{\Phi_1(v,j),\Phi_2(v,j),\Phi_3(v,j)\}\in inf(v, j)$. Moreover, it is unknown whether this grouping is unique.

\subsection{Edge equality analysis}
Although the update function of 3-WL is different from that of 3-FWL, the method of identifying common edges introduced in Section \ref{sec: Identify common edges.} is still applicable. By looking for tuples belonging to class 2) in $\{\{(c^t_{\Phi_1(v,j)})|j\in[N]\}\}$, common edge $CE(\Phi_1(v,j), v)$ can be determined. Similarly, $CE(\Phi_2(v,j), v)$ and $CE(\Phi_3(v,j), v)$ can also be obtained. 

Apply the method similar to that in Section \ref{sec: turn-over trick.}, remove $CE(\Phi_1(v,j), v)$ in $c^t_{\Phi_1(v,j)}$, and get $(NE_{j,1}, NE_{j,2})$ for every $j$. Similarly, get $(NE_{j,3}, NE_{j,4})$ and $(NE_{j,5}, NE_{j,6})$ from $c^t_{\Phi_2(v,j)}$ and $c^t_{\Phi_3(v,j)}$, respectively. 

The process of grouping is to take out three tuples from $\{\{(NE_{j,1}, NE_{j,2})|j\in [N]\}\}$, $\{\{(NE_{j,3}, NE_{j,4})|j\in [N]\}\}$, $\{\{(NE_{j,5}, NE_{j,6})|j\in [N]\}\}$ respectively. Moreover, these three tuples must satisfy edges equality conditions. And edge equality is transitive, which makes the analysis process extremely cumbersome. We use the following Figure \ref{fig: Edge equality analysis.} for edge equality analysis. 

\begin{figure}[htbp]
  \centering
  {\includegraphics[width=3cm]{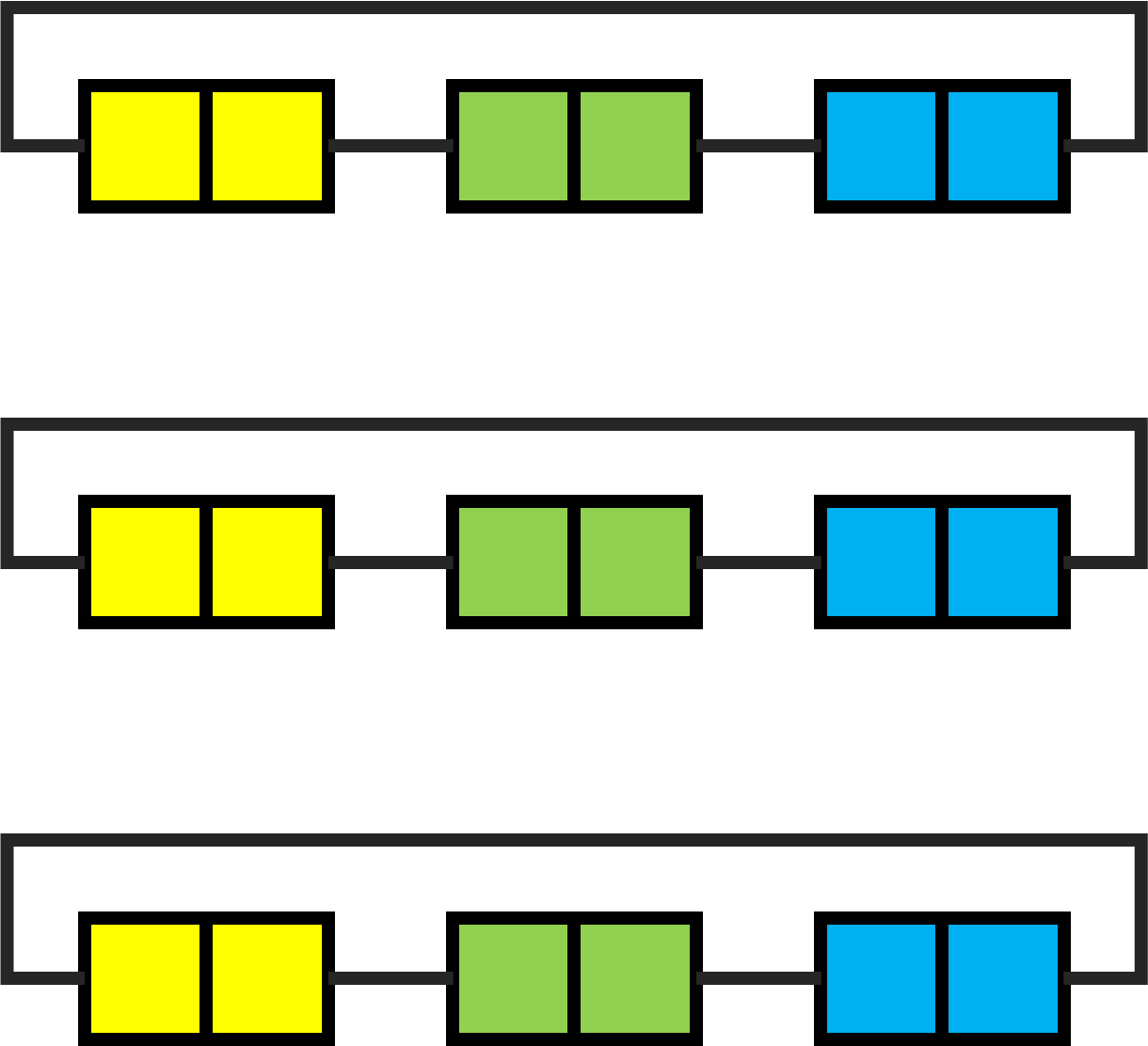}}
  \label{fig: Edge equality analysis.}
  \caption{Edge equality analysis.}
\end{figure}

Each tuples in $\{\{(NE_{j,i}, NE_{j,i+1})|j\in [N]\}\},i=1,3,5$ is represented by a $1\times2$ rectangular block, that is, each square in Figure \ref{fig: Edge equality analysis.} represents a $NE$. The yellow ones represent the tuples in $\{\{(NE_{j,1}, NE_{j,2})|j\in [N]\}\}$, the green ones represent the tuples in $\{\{(NE_{j,3}, NE_{j,4})|j\in [N]\}\}$, and the blue ones represent the two-tuples in $\{\{(NE_{j,5}, NE_{j,6})|j\in [N]\}\}$. The two blocks connected by the black line represent that their corresponding $NE$s are spliced together, so the lengths of the edges corresponding to the blocks connected by the black line are the same (corresponding to the same flute in the tetrahedron). Figure \ref{fig: Edge equality analysis.} shows the real grouping situation, that is, the tuples in each row come from the same $j$.

We reformulate the tuples grouping problem as the following problem: Is there another matching (grouping) way that can produce new "tetrahedrons"? We use the following two examples to illustrate how to use the "edge equality analysis" for this problem.

\begin{figure}[htbp]
  \centering
  {\includegraphics[width=7cm]{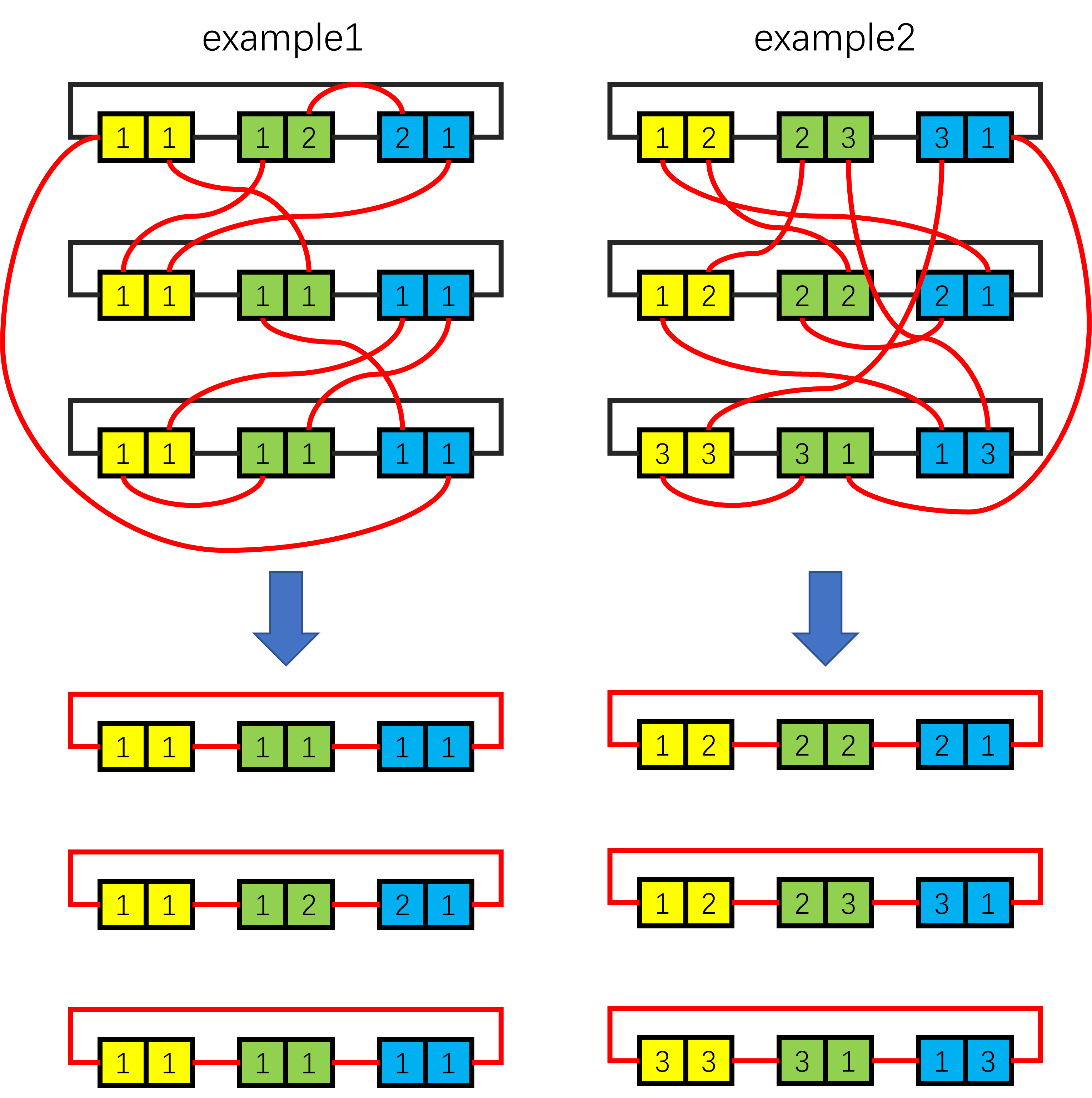}}
  \label{fig: Two edge equality analysis examples.}
  \caption{Two edge equality analysis examples.}
\end{figure}

First, select a yellow rectangle, a green rectangle, and a blue rectangle (the reason for each one is that they are to be spliced with the three edges of the root triplet), and connect them with a red line. The meaning of the red line and the black line are the same, that is, it represents the $NE$ splicing corresponding to the two squares. Therefore, the corresponding edges' lengths of the blocks connected by the red lines are also equal. After connecting all the squares with the red line, it means that a new grouping has been created.

Then analyze which edges are of equal length. For any unnumbered square, assign it an unused number, such as 1. Subsequently, since all the edges joined to this edge have the same length, the squares connected to this square (including black lines or red lines) are marked with the same number. Since equal edge lengths are transitive, all squares connected to a numbered square must be marked with the same number.

Repeat the above labeling process until all squares have labels. Squares with the same number indicate that their corresponding $NE$s are of equal length. Then, remove the black line, and expand all the rectangles into the structure shown in Figure \ref{fig: Edge equality analysis.} according to the connection of the red line. By comparison, in the two examples in Figure \ref{fig: Two edge equality analysis examples.}, no new tetrahedron is generated.

In fact, the new grouping (the grouping represented by the red line) can be changed from the real grouping (the grouping represented by the black line) in two ways: 1) exchange tricks, 2) turn-over tricks. Through these two tricks, the grouping shown by the black line can be changed to the grouping shown by the red line. Although many possible groupings can be produced by these two tricks, at the same time, each time these tricks are used, a new constraint will be added, that is, the edge equality constraint. As can be seen from Figure \ref{fig: Two edge equality analysis examples.}, due to the existence of black and red lines and the transitivity of edge equality, many edges must be the same. In other words, both exchange tricks and turn-over tricks are at the expense of "edge length freedom".

\subsection{Future work}
So far, we have not found an effective way to prove that 3-WL can distinguish all 3D graphs, nor have we found a counterexample. We will use the "edge equality analysis" method to analyze whether 3-WL is sufficient to solve all 3D graphs isomorphism problem.

\section*{References}
\medskip
{
\small

[1] Li Z, Wang X, Huang Y, et al. Is Distance Matrix Enough for Geometric Deep Learning?[J]. arXiv preprint arXiv:2302.05743, 2023.

[2] Morris C, Ritzert M, Fey M, et al. Weisfeiler and leman go neural: Higher-order graph neural networks[C]//Proceedings of the AAAI conference on artificial intelligence. 2019, 33(01): 4602-4609.

[3] Huang N T, Villar S. A short tutorial on the weisfeiler-lehman test and its variants[C]//ICASSP 2021-2021 IEEE International Conference on Acoustics, Speech and Signal Processing (ICASSP). IEEE, 2021: 8533-8537.

[4] Shervashidze N, Schweitzer P, Van Leeuwen E J, et al. Weisfeiler-lehman graph kernels[J]. Journal of Machine Learning Research, 2011, 12(9).

[5] Cai J Y, Fürer M, Immerman N. An optimal lower bound on the number of variables for graph identifications[J]. Combinatorica, 1992, 12(4): 389-410.


\end{document}